\documentclass[onecolumn,showpacs,preprintnumbers,amsmath,amssymb,footinbib]{revtex4-1}
\usepackage[utf8]{inputenc}
\usepackage[english]{babel}
\usepackage{lmodern}
\usepackage{amsmath}
\usepackage{graphicx}
\usepackage{amssymb}
\usepackage{epstopdf}
\usepackage{ upgreek }
\usepackage{color}
\usepackage{setspace}
\usepackage{hyperref}

\graphicspath{{Figures/}}

\usepackage[toc,page]{appendix}
\usepackage{tocloft}

\begin{document}

\title[]{Masked states of an atom coupled to a standing-wave cavity mode}

\author{R.~Guti\'{e}rrez-J\'{a}uregui}
\email[Email:]{r.gutierrez.jauregui@gmail.com}
\affiliation{Department of Physics, Columbia University, New York, NY, USA.}

\date{}

\begin{abstract}

The form of the eigenstates of an atom coupled to a cavity mode displaying a three dimensional periodic profile are obtained. It is shown that the quantized motion leads to degenerate states where the atomic degrees of freedom are masked, that is, upon detection of one component of this composite system the others remain in an entangled state. When the system is extended to include drive and dissipation it is found to undergo a dissipative quantum phase transition at a critical drive amplitude. Unlike other phase transitions reported in the literature, the degeneracy prepares the system in a superposition of incompatible states upon detection of the electromagnetic field. Probing the field hints at an order above the transition point that, due to state masking, allows for atomic coherence to survive at long times.

\end{abstract}

\maketitle

%%%%%%%%%%%%%%%%%%%%%%%%%%
\section{Introduction}  %%
%%%%%%%%%%%%%%%%%%%%%%%%%% 
The first experimental evidence of the discrete quantum nature of the electromagnetic field represents one of the crowning achievements of cavity quantum electrodynamics and provides a beautiful demonstration of measurements at the quantum level~\cite{Brune_1996}. The demonstration relied on the coherent interaction between the microwave photons stored inside a superconducting cavity and a traversing atom. This interaction allowed for the atom to act as a probe---a secondary system coupled to the state of the field---that acquired information as correlations built up. Since the cavity minimizes the coupling to the surrounding environment, detection of the atomic state at different interaction times was used to infer the photon number distribution of the field. Given the extraordinary control acquired over quantum systems, it has become possible to exploit other correlations between light and matter to test fundamental principles~\cite{Haroche_1991,Castin_1991,Wilkens_1992a,Meekhof_1996}. For example, the correlations that rise between the dynamical variables of light and matter when a structured beam is shinned upon an atomic cloud have been used to detect the orbital angular momentum of the photon~\cite{Andersen_2006}. And those that rise between internal and external degrees of freedom of a tightly trapped ion to probe abrupt changes in the spatial distribution of a structured field~\cite{Jauregui_2004,Schmiegelow_2012,Schmiegelow_2016}.

Recently, this control has been extended to driven-dissipative settings where the interaction between light and matter is tailored through external fields~\cite{Dimer_2007, Baumann_2010,Leonard_2017}. As a result, these quantum open systems can be used to access a regime where dissipation and coherence compete to realize novel quantum states~\cite{Domokos_2002, Nagy_2010,Baumann_2011,Carmichael_2015,Kurizki_2016,Fink_2017,Bottcher_2019}. These states can still be probed from the information leaked to the environment following the same principles as above. Take the breakdown of photon blockade as an example~\cite{Carmichael_2015,Fink_2017}. It considers an extension of the two-state atom coupled to a cavity-mode where the mode is now driven by an external coherent field. The system displays two different behaviors depending on the strength of the driving field amplitude. For weak driving fields, the absorption of one photon shifts the system out of resonance with the drive, leading to a maximum of a single photon inside the cavity and therefore photon antibunching in the cavity output channel. After the driving field is ramped up past a limiting value, however, the cavity mode populates and output photon statistics display coherence. The phase of the output photons is associated to one of two possible atomic polarizations and its measurement allows for the state of the atom to be infered~\cite{Alsing_1991}. Similar behavior is encountered in the case of atomic self-organization when the interaction between the external degrees of freedom and a standing-wave cavity mode is considered~\cite{Black_2003,Domokos_2002, Nagy_2010,Baumann_2011}. The abrupt changes in behavior found in both cases can be studied in analogy to quantum phase transitions in thermal equilibrium where the system undergoes an order-disorder transition, an analogy that can be traced back to the early days of the laser~\cite{Graham_1970,Bonifacio_1978,Rice_1994}.

This paper concerns the correlations between a single atom coupled to a single mode of a standing-wave cavity. It is built around the multipartite entanglement that correlates photon number, atomic momentum, and electronic state. The effect of this entanglement in the context of quantum phase transitions is rarely discussed in the existing literature. Usual models are based on physical considerations from which one degree of freedom can be parametrized and then focus on the correlations between the remaining two, whether it is the effect of the atomic position as an effective coupling strength~\cite{Brune_1992,Fink_2017}, the cavity field mediating the interaction between internal and external atomic degrees of freedom~\cite{Lin_2011,Dalibard_2011}, or the internal field creating a bridge for the field to connect center-of-mass states~\cite{Domokos_2002a,Black_2003,Baumann_2010}. When the three components are accounted for, new conserved operators emerge and the model Hamiltonian displays degeneracies. In particular, when each degree of freedom can be found in one of two states (states $\vert \uparrow_{a} \rangle$ and $\vert \downarrow_{a} \rangle$ for $a=1,2,3$ degrees of freedom) the combined ground state is of the form (Sec.~\ref{Sec:structure}) $$\vert \Psi \rangle = \frac{\vert \uparrow_{1}, \uparrow_{2}, \uparrow_{3} \rangle + \vert \uparrow_{1}, \downarrow_{2}, \downarrow_{3} \rangle + \vert \downarrow_{1}, \uparrow_{2}, \downarrow_{3} \rangle + \vert \downarrow_{1}, \downarrow_{2}, \uparrow_{3} \rangle}{2} . $$ It can be thought of as if any two degrees of freedom conspire to create an effective state for the remaining degree to probe. The atomic degrees of freedom mask one another so the field can not be used to determine their individual states.

Atomic state masking translates to the driven dissipative scenario where a lossy standing-wave cavity is driven by an external coherent drive. Here, the system undergoes a phase transition above a critical driving field amplitude like the breakdown of photon blockade or self-organization~(Sec.~\ref{Sec:mean-field}). Ordering, however, does not assure a formation of regular patterns in the traditional sense due to masking. The order now defines a subspace for coherent dynamics to unfold since detection of output photons leaves the system in an entangled state. The system reaches order above the critical drive, but lack of information accessible to the probing field allows for quantum correlations between atomic degrees of freedom to survive. 

The structure of the paper is as follows. In Sec.~\ref{Sec:structure} the model Hamiltonian is presented in a way that draws attention towards the correlations that follow the exchange of excitation and momentum. Building upon previous results by Ren and Carmichael~\cite{Ren_1992,Ren_1995} these correlations are shown to cause degeneracies that obscure the information available to an external observer probing any single degree of freedom. To explore the impact of this result, a particular observer is introduced in Sec.~\ref{Sec:open_system} when the effect of a surrounding environment is considered. This observer has access to photons that leak through the cavity walls and is able to control the photon input through an external coherent drive. The resulting system is shown to display two competing behaviors as the parameter space is explored. This result is supported by both a mean-field study and a numerical evolution accounting for quantum correlations in Secs.~\ref{Sec:mean-field} and~\ref{Sec:ordering}, respectively. The mean field results predict ordered solutions, but, being dynamically unstable throughout the parameter space, they appear to be unreachable. It is until quantum correlations are taken into account using a master equation evolution that order is encountered. The results point towards the importance of quantum correlations in this composite quantum system and, unlike traditional results, how extra degrees of freedom can lead to coherence. A summary of the work and conclusions is given in Sec.~\ref{Sec:conclusion}.

%%%%%%%%%%%%%%%%%%%%%%%%%%%%%%%%%%%%%%%%%%%%%%%%%%%%%%%%%%%%%%%%%%%%%%%
\section{A two-state atom inside a cavity}\label{Sec:structure} %%
%%%%%%%%%%%%%%%%%%%%%%%%%%%%%%%%%%%%%%%%%%%%%%%%%%%%%%%%%%%%%%%%%%%%%%%
 
In this work I consider the case of a two-state system of mass $M$ and momentum $\hat{\mathbf{p}}$ interacting with a single cavity mode. When the conditions for dipolar coupling and rotating-wave approximations are met~\cite{Carmichael_1999,Babiker_2002, Haroche_2006}, their evolution is described by the Hamiltonian
\begin{equation}\label{eq:coup_1}
\hat{\mathcal{H}} = \frac{\hat{\mathbf{p}}^{2}}{2M} + \hbar \omega_{c} \hat{a}^{\dagger} \hat{a} + \hbar \omega_{a} \hat{\sigma}_{+}\hat{\sigma}_{-} + \mathbf{d}\cdot\mathbf{E}\left(\mathbf{x} \right) \left( \hat{a} \hat{\sigma}_{+} + \hat{a}^{\dagger} \hat{\sigma}_{-} \right) \, ,
\end{equation}
with $\hat{a}$ and $\hat{a}^{\dagger}$ photon annihilation and creation operators for the cavity mode; $\hat{\sigma}_{+}$ and $\hat{\sigma}_{-}$ raising and lowering operators for the two-state system that satisfy the commutation relations $$[\hat{\sigma}_{+},\hat{\sigma}_{-}]= \hat{\sigma}_{3} \, , \, [\hat{\sigma}_{\pm}, \hat{\sigma}_{3}]  = \mp 2 \hat{\sigma}_{\pm};$$ and parameters $\omega_{a}$, $\omega_{c}$ to represent the resonance frequency of the two-level transition and field mode frequency. The first three terms of Eq.~(\ref{eq:coup_1}) describe the free evolution of the cavity mode and two-state system while the last term describes their coupling. This coupling depends on the atomic position through the local field amplitude $\mathbf{E}(\mathbf{x})$ projected along the electric dipole axis $\mathbf{d}$. And, as such, it allows for the atom to probe the spatial profile of the mode. Throughout this section the coupling is considered to be 
\begin{equation}\label{eq:coup_2}
\mathbf{d}\cdot\mathbf{E}(\mathbf{x}) = \hbar \Omega \sin( k_{1}x_{1})\sin( k_{2}x_{2})\cos( k_{3}x_{3}) \, , 
\end{equation}
as found in the idealized scenario of a polarized atom coupled to a standing-wave cavity in three dimensions. 

The standing-wave cavity provides a useful setting where the correlations among internal and external degrees of freedom can be studied.  Its normal modes are characterized by a wavevector $\mathbf{k}$ with $k_{m}$ components along the $x_{m}$-directions $(m=1,2,3)$ and their periodic profile gives way to physical interpretation through simple algebraic manipulation. The analysis that follows, however, is readily extended to different cavity geometries, \text{e.g.} cylindrical cavities used in early maser experiments~\cite{Meschede_1985,Jauregui_2005}, open-ended parabolic cavities~\cite{Maiwald_2009,Jauregui_2018}, or even rectangular cavities where one wall is removed to allow for external manipulation of the atomic state~\cite{Baumann_2010}. With the model Hamiltonian~(\ref{eq:coup_1}) describing the elementary process of absorption and emission of photons by the two-state system as it transitions among excited $\vert e \rangle$ and ground $\vert g \rangle$ states, the spatial structure simply determines the rate at which this process occurs. The rate is maximized when the spatial distributions of the external states $\vert \phi^{(e)}\rangle$ and $\vert \phi^{(g)}\rangle$ match that of the field. This condition is fulfilled for an otherwise free atom since the external states follow the symmetries imposed by the surrounding cavity mode and, to some extent, mimic its spatial profile. In other words, the dynamical variables of the field become candidates to describe the atomic center-of-mass. For the standing-wave cavity this means that external states can be expanded as a linear combination of plane-waves that are both eigenstates of the momentum operator,
\begin{equation}
\hat{p}_{m} \vert  q_{m} +l_{m} k_{m} \rangle  = \hbar (q_{m} + l_{m} k_{m}) \vert  q_{m} + l_{m} k_{m} \rangle \, , 
\end{equation}
and reflect the periodicity of the cavity mode with $l_{m} = 0, \pm 1, \pm 2, \dots $ and quasimomentum $q_{m} \in [-k_{m}/2 , k_{m}/2)$. 

The previous description merely sets a natural basis for the unbound states of atom and cavity mode. Entanglement enters the picture through the exchange of excitations and momenta. Conservation of excitations, on the one hand, cause the eigenstates of Eq.~(\ref{eq:coup_1}) to take the form
\begin{align}\label{eq:centre_2}
&\vert \Psi \rangle = \vert \phi^{(e)} \rangle \vert e, n-1 \rangle  +  \vert \phi^{(g)} \rangle \vert g, n \rangle  \, ,
\end{align} 
for a cavity containing $n=1,2,\dots$ photons, thus organizing the dynamic into Jaynes-Cummings-like doublets plus a single unpaired ground state $\vert\phi^{o}\rangle\vert g,0\rangle$. Conservation of momentum, on the other hand, is represented by the operators~\cite{Ren_1995,Larson_2008}
\begin{equation}
\hat{\mathcal{O}}_{m} = \exp \left( \frac{i \pi \hat{p}_{m}}{\hbar k_{m}}  \right)  \hat{\sigma}_{3} \, , \label{eq:parity_z} 
\end{equation} 
and causes the eigenstates to organize into parity chains that connect inner and external states along each momentum component:  
\begin{align}
&\cdots \leftrightarrow \vert 0,e,n-1 \rangle  \leftrightarrow \vert k_{m},g,n \rangle  \leftrightarrow \vert 2k_{m},e,n-1 \rangle \leftrightarrow \cdots \, (\eta_{m} = +1) \label{eq:parity_1}\\
&\cdots \leftrightarrow \vert 0,g,n \rangle  \leftrightarrow \vert k_{m},e,n-1 \rangle  \leftrightarrow \vert 2k_{m},g,n \rangle \leftrightarrow \cdots \, \, \, \, \, \, \, \, \, \, \,  (\eta_{m} = -1)\label{eq:parity_2}
\end{align} 
Here, $\eta_{m} = \pm 1$ correspond to the eigenvalues of $\hat{\mathcal{O}}_{m}$ and the quasimomentum $q_{m}$ has been removed from the notation since it remains constant. From Eqs.~(\ref{eq:parity_1}) and~(\ref{eq:parity_2}) the exchange of interactions that correlate photon number and electronic state are seen to be accompanied by an exchange of momentum that correlates electronic states and center-of-mass motion. The dynamics then follow standard Rabi oscillations with a caveat: with each absorption or emission of a photon the center-of-mass spreads over the momentum ladder in steps of $\vert \hbar k_{m} \vert$ length. The possibility to move up and down the ladder is attributed to the structure of the cavity mode itself. Being formed by a superposition of counter-propagating waves, the mode presents the atom with two paths to follow at each half Rabi cycle (for each component $\pm k_{m}$). The ambiguity created by this choice of path is reflected in the eigenstates of $\hat{\mathcal{H}}$ which, unlike excitations, are restricted to manifolds of given parity rather than given total momentum (cavity mode plus atom). 

There is an apparent similarity between the parity chains and the eigenstates of the Rabi Hamiltonian~\cite{Casanova_2010,Braak_2011}. The similarity arrives from the inclusion of both rotating and counter-rotating terms in the Rabi model that play the role of the standing-wave in the current model. It is then expected for the eigenstates of the Rabi Hamiltonian and those of the $\hat{\mathcal{O}}_{m}$ operators to share the same structure. The former is found inside a photon-number ladder and the latter by an equivalent momentum one. The main difference between the two models resides in the length of the steps of each ladder. While the free-mode energy spectrum displays a harmonic structure the kinetic energy is anharmonic, meaning that the momentum states can be Doppler shifted out of resonance. This shift can be neglected for light-matter systems displaying small recoil energies, allowing for external degrees of freedom to be used to simulate effects originally conceived for inner degrees. The similarity also allows for results obtained under the Jaynes-Cummings and Rabi models to be exploited for the current one. For example, if instead of a standing-wave cavity, a waveguide with defined traveling mode was considered, then the eigenstates of the Hamiltonian~(\ref{eq:coup_1}) would be restricted to manifolds of total momentum $\hat{P} = \hbar k_{m} \hat{\sigma}_{3} + \hat{p}_{m}$. In this case knowledge of one component defines unequivocally the others and the system can be probed by measuring internal or external degrees of freedom.

In the remainder of this work I consider the cavity mode to be tuned in exact resonance with the atomic transition, $\omega_{a} = \omega_{c}$. The parity chains become degenerate under this condition and the model Hamiltonian is diagonalized by the upper $(+)$ and lower $(-)$ dressed states $$\vert \pm , n \rangle = \frac{\vert e, n-1 \rangle \pm \vert g,n \rangle }{\sqrt{2}}. $$ Equation~(\ref{eq:coup_1}) takes the form
\begin{equation}\label{eq:on-resonance}
\hat{\mathcal{H}} =  \frac{\hat{\mathbf{p}}^{2}}{2M}  + \hbar \omega_{a} \hat{n} + \sum_{n} \sqrt{n} \mathbf{d}\cdot\mathbf{E}\left(\mathbf{x} \right)  \left[ \vert + ,n \rangle \langle +,n \vert - \vert -,n \rangle \langle -,n \vert \right]  \, ,
\end{equation}
where the center-of-mass motion is conditioned on the internal state. The atom remains sensitive to the phase and amplitude of the field, but the dressed states provide effective potentials for the dynamics to unfold. Entanglement between internal and external degrees of freedom develops from these two possible paths.

%%%%%%%%%%%%%%%%%%%%%%%%%%%%%%%%%%%%%%%%%%%%%%%%
\subsection{Dynamics: which-path information}   %%%
%%%%%%%%%%%%%%%%%%%%%%%%%%%%%%%%%%%%%%%%%%%%%%%%

As a prelude to state-masking presented in the following subsection consider the evolution of a two-state system prepared with a spread in momentum that is significantly smaller than the photon momentum ($\Delta p_{m} \ll \hbar k_{m}$). A two-state system prepared this way is equipped with a resolution that allows it to probe the spatial profile of the field. This set-up has been exploited in matter wave diffraction experiments where, traditionally, the superposition of two counterpropagating lasers create a grating for traversing atoms to be diffracted on~\cite{Gould_1986,Kunze_1996,Peik_1997}. The same dynamics follow when the field is quantum in nature. Here the internal states and cavity mode correlate to create two possible gratings determined by upper and lower branches in Eq.~(\ref{eq:on-resonance}). As the atom traverses the cavity the composite system evolves into the state~\cite{Ren_1995}
\begin{equation}\label{eq:dynamical_1}
\vert \psi(t) \rangle \simeq e^{i\omega_{\text{\tiny{D}}}t} \left( c^{(+)} \vert \psi^{(+)}(t) \rangle + c^{(-)} \vert \psi^{(-)}(t) \rangle \right)  \, ,
\end{equation}
where the Doppler shift is approximated by a central value $\omega_{\text{\tiny{D}}}$ leading to an overall phase (valid for small recoil frequencies and short times) and the conditional states are
\begin{equation}\label{eq:dynamical_2}
\vert \psi^{(\pm)}(t) \rangle =  \exp \left[\pm \frac{i \sqrt{n}t}{\hbar}\mathbf{d}\cdot\mathbf{E}\left(\mathbf{x} \right)  \right]  \vert \phi(t_0), \pm ,n  \rangle \, .
\end{equation}
The conditional states differ by a relative phase that causes the atomic motion to carry information of the mode photon number $n$ in addition to that of its spatial profile. In a similar fashion, by considering more spatial dimensions, each degree of freedom correlates with others; adding to the information embedded into each degree-of-freedom and allowing for momentum components to probe one another. This is better shown by writing the dipole potential~(\ref{eq:coup_2}) in the form
\begin{equation}\label{eq:dynamical_3}
\mathbf{d}\cdot\mathbf{E}(\mathbf{x}) = \frac{-\hbar \Omega}{4} \sum_{j,j^{\prime}=0,1} (-1)^{j} \cos[k_{1}x_{1}+(-1)^{j}k_{2}x_{2}+(-1)^{j^{\prime}}k_{3}x_{3}] \, ,
\end{equation}
displaying four terms that share the same structure but differ in phase. Each term corresponds to a particular combination of momentum components. For example, when one wall of the cavity is removed the two-dimensional coupling
\begin{equation}\label{eq:two_dimension}
\mathbf{d} \cdot \mathbf{E}(x) \big\rvert_{2D} = \hbar \Omega \cos k_1 x_1 \sin k_{2} x_{2} \, 
\end{equation}
can be written as
\begin{equation}\label{eq:two_dimension1}
2 \cos k_1 x_1 \sin k_{2} x_{2} = \sin [k_1 x_1 - k_{2} x_{2}] - \sin [k_1 x_1 + k_{2} x_{2}] \, .
\end{equation}
These two terms act upon the correlated ($\hat{p}_{1}+\hat{p}_{2}$) and anti-correlated ($\hat{p}_{1}-\hat{p}_{2}$) momentum components. The operators commute with each other and form a complete basis in two dimensions, which can then be used to diagonalize the Hamiltonian in the same way as the dressed states above. 

\begin{figure}[h]
\begin{center}
\includegraphics[width=.8\linewidth]{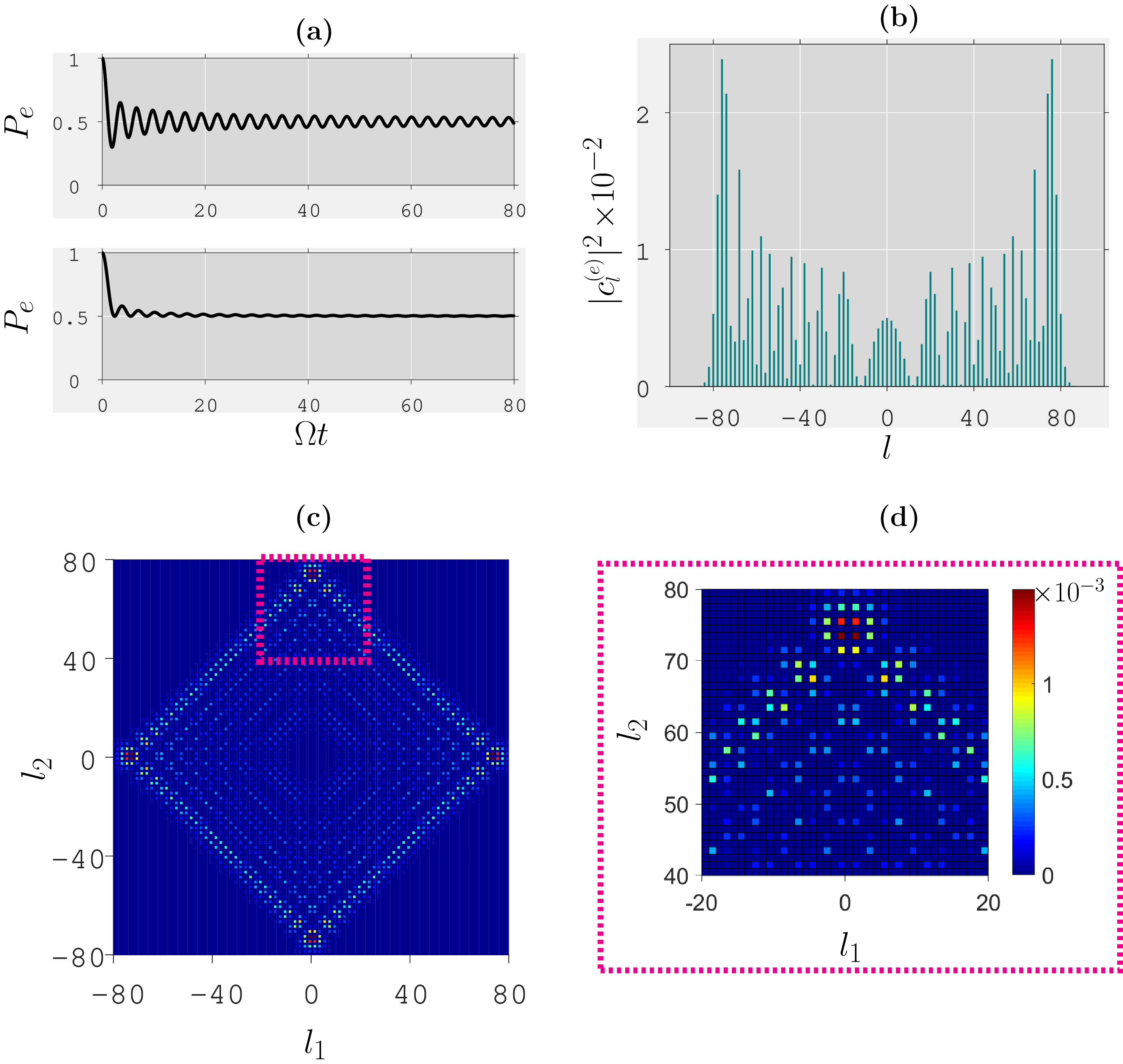}
\caption{Quantum Rabi oscillation signal and momentum distribution for $\omega_{\text{r}} = 10^{-4}\Omega$. (a) Probability $P_{e}$ of finding the atom in the excited state as a function of time for one (up) and two dimensions (down). (b)-(d) Probability distribution $\vert c_{l_{m}}^{(e)}\vert^{2}$ of finding the atom in an excited state with momentum components $l_{m}k_{m}$ at time $\Omega t = 80$ for one (b) and two dimensions (c), (d). Only states of momentum $l_{m}k_{m}$ with even $l_{m}$ are found. }\label{figure_1D}
\end{center}
\end{figure}

For each additional dimension the atomic state is presented with more paths---or gratings---to follow. The outgoing state reflects these correlations and can be obtained by using Eqs.~(\ref{eq:dynamical_2}) and (\ref{eq:two_dimension}) and the Jacobi-Anger expansion
\begin{equation}\label{eq:jacobi_anger}
\exp[i \Omega t  \cos k x] = \sum_{m} i^{m} J_{m}(\Omega t) e^{i m k x} \, ,
\end{equation}
where $J_{m}$ is the Bessel function of order $m$~\cite{Abramowitz_1972}. When the four terms of Eq.~(\ref{eq:dynamical_3}) are taken into account this leads to a product of Bessel functions and exponentials that determine the spread in momentum distribution. Take the case of a one-dimensional potential where the conditioned states are
\begin{align}
\vert \psi^{(\pm)}(t) \rangle &= \sum_{n,m}  J_{m}(\sqrt{n} \Omega t) (\pm e^{i kx})^{m} \vert  \phi(t_0), \pm ,n  \rangle \, .
\end{align}
The relative phase causes a drift between the center-of-mass states in momentum space that allows them to be distinguished with better resolution as time advances. With less and less overlap among the conditional external states a suppression of interference follows. A first mark of this distinction is found in the Rabi oscillations of an atom initially prepared in the excited state with low momentum $(l\simeq 0)$ traversing an empty cavity. The Rabi oscillations dampen due to the loss of coherence as the overlap between conditional states decreases. For one dimension the damping is given by
\begin{equation}
\langle \phi^{(+)}(t) \vert \phi^{(-)}(t) \rangle \big\rvert_{1D} = \sum_{m} (-1)^{m} \vert J_{m}(\Omega t) \vert^{2} \, 
\end{equation} 
while for two dimensions this rate increases to 
\begin{equation}\label{eq:random_2D}
\langle \phi^{(+)}(t) \vert \phi^{(-)}(t) \rangle \big\rvert_{2D} = \sum_{m_{1},m_{2}} (-1)^{m_{1}+m_{2}} \left\vert J_{m_{1}} \left(\frac{\Omega t}{2} \right) J_{m_{2}}\left(\frac{\Omega t}{2}\right) \right\vert^{2} \, .
\end{equation}
As the dimension of the Hilbert space increases so does the number of possible paths the atom can explore leading to a higher damping rate. In Figure~\ref{figure_1D}(a) the damped Rabi oscillations are presented for the one and two-dimensional potentials for comparison. The atom is originally prepared in the excited state with a defined momentum ($l_{m}=0$) inside an empty cavity.

The loss of coherence has an impact on the atomic system that goes beyond the damping of the Rabi oscillations. As described above, at each half-Rabi cycle the atom exchanges an excitation with the mode that is accompanied by a displacement in each momentum component. The displacement occurs in two directions $(\pm k_{m})$ with equal probability amplitudes and leaves the atomic momentum in a superposition state. After many cycles have passed these choices cause the momentum distribution to have traced a quantum random walk~\cite{Cohen_1992,Aharonov_1993,Kempe_2003}. This is displayed in Figure~\ref{figure_1D}(b) where the distribution is presented for a one dimensional cavity $(\mathbf{d} \cdot \mathbf{E}(x) = \hbar \Omega \cos k x )$ at a time $t=80 \Omega^{-1}$, long before the Doppler shift becomes comparable to $\Omega$. The two-peaked distribution is characteristic of the quantum random walk as explored theoretically for optical~\cite{Bouwmeester_1999,Agarwal_2005} and cold atom systems~\cite{Summy_2016}. The situation is richer when more spatial dimensions are considered. Figures \ref{figure_1D}(c) and \ref{figure_1D}(d) display the two-dimensional case, where the random walk is performed over the correlated and anti-correlated momentum components in accordance with Eq.~(\ref{eq:two_dimension}). When the average over one momentum component is taken, the two-peaked distribution gives way to a single-peaked one Gaussian distribution. This distribution is characteristic of the classical random walk.

%%%%%%%%%%%%%%%%%%%%%%%%%%%%%%%%%%%%%%%%%%%%%%%%%%%%%%%%%%%%%%%%%%%%%%%%%%%%%%
\subsection{Masked states}\label{Sec:entanglement}
%%%%%%%%%%%%%%%%%%%%%%%%%%%%%%%%%%%%%%%%%%%%%%%%%%%%%%%%%%%%%%%%%%%%%%%%%%%%%%

The damping of Rabi oscillations and coherences that gave rise to the quantum random walk provide two paradigmatic examples of decoherence. Information acquired by a probe can destroy quantum coherences and veil effects that we have come to associate with the quantum. In the approach taken above, the probe consisted of one momentum component that correlated to internal and external degrees of freedom of the system following a standard which-path scenario. Different sets of paths were introduced systematically from the pairing between system components, \textit{e.g.}, two-state energy and photon number [see Eq.~(\ref{eq:on-resonance})] or momentum correlations [Eq.~(\ref{eq:dynamical_3})]. Since the system was monitored the distinguishability of the paths was the main source of decoherence: each pair of paths differed from one another through a relative phase, a binary parameter that could take the values zero or $\pi$. For couplings of the form $$\hbar \Omega \sin( k_{1}x_{1})\sin( k_{2}x_{2})\cos( k_{3}x_{3})(\hat{a} \hat{\sigma}_{+} + \hat{a}^{\dagger} \hat{\sigma}_{-}), $$ these phases can cancel out and a degeneracy of the Hamiltonian is expected. This degeneracy obscures the information that can be obtained by probing any one component as shown in the following.

\begin{figure}[h]
\begin{center}
\includegraphics[width=.4\linewidth]{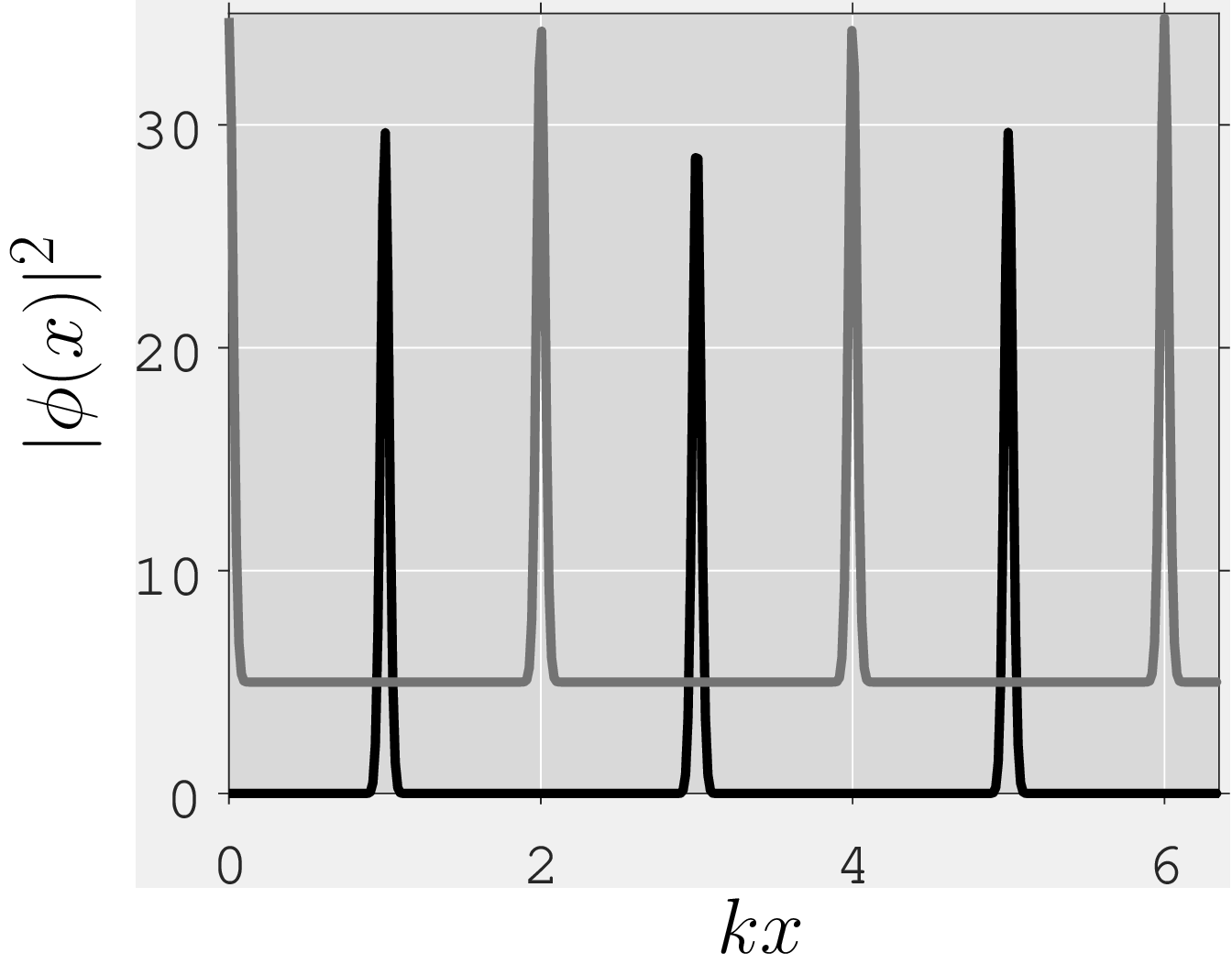}
\caption{Lowest energy center-of-mass states conditioned to two effective potentials due to the internal states for $\omega_{\text{r}} = 10^{-4}\Omega$ and $n=1$. Black (gray) curves indicate upper (lower) dressed-states. The gray curve is displaced vertically by five for clarity. }\label{figure_2D}
\end{center}
\end{figure}

I now shift the attention from the system dynamics to the lowest non-trivial eigenstate of the system, but remain in a regime where the coupling strength dominates over the recoil frequency. In this regime an atom inside a standing-wave cavity tends to localize such that fluctuations in position are minimized. This localization occurs around the minima of the effective potential, a region that is conditioned to the internal state. In one dimension and for a system found along the upper dressed-state branch the atom localizes around $k x = 2n\pi$ (denoted by a state $\vert x_{\uparrow} \rangle$) and when found along the lower branch it localizes around $k x = (2n+1)\pi$ (state $\vert x_{\downarrow}\rangle$). The lowest energy state is composed of a superposition of these two degenerate states
\begin{equation}\label{eq:localization_mask}
\vert \phi_{1D} \rangle = \frac{  e^{i \varphi}\vert +,1 \rangle \vert x_{\uparrow} \rangle + e^{-i \varphi}\vert -,1 \rangle \vert x_{\downarrow} \rangle}{\sqrt{2}} \, .
\end{equation}
Equation~(\ref{eq:localization_mask}) establishes that, for conditions of strong localization in one dimension, knowledge of the external state determines the internal state and vice-versa. The two state components are strongly entangled and, as such, are good probes for one another. This is exemplified in Figure~\ref{figure_2D} where the conditioned external states are plotted in the position representation for the same set of parameters as in Figure~\ref{figure_1D}. The states do not overlap for this choice of parameters. The phase $\varphi$ carries information regarding the population of the system on each parity branch and is set by the initial conditions. 

The situation changes when more dimensions are taken into account since the region of localization is also conditioned to the correlations in momentum. For two dimensions the lowest energy state becomes
\begin{align}\label{eq:stationary_2}
\vert \phi_{2D} \rangle &= \frac{\vert +,1 \rangle \left[ e^{i (\varphi_{1}+\varphi_{2})} \vert x_{1\uparrow}, x_{2\uparrow} \rangle + e^{-i (\varphi_{1}+\varphi_{2})}\vert x_{1\downarrow}, x_{2\downarrow} \rangle\right]}{2} \nonumber \\
&+\frac{ \vert - , 1 \rangle \left[ e^{i (\varphi_{1}-\varphi_{2})}\vert x_{1\uparrow}, x_{2\downarrow} \rangle + e^{-i (\varphi_{1}-\varphi_{2})}\vert x_{1\downarrow}, x_{2\uparrow} \rangle\right]}{2}  \, 
\end{align}
as expected from the potential~(\ref{eq:two_dimension1}). The eigenstates describe a checkerboard pattern with alternating regions of localization. The existence of only two regions is caused by the dichotomy introduced by each degree of freedom and has deeper consequences: since each state can only accommodate one of two values its detection conditions the remaining components to a superposition. For example, by projecting the state over the upper branch
\begin{align}\label{eq:stationary_2}
\langle + ,1 \vert \phi_{2D} \rangle &=  \frac{e^{i (\varphi_{1}+\varphi_{2})} \vert x_{1\uparrow}, x_{2\uparrow} \rangle + e^{-i (\varphi_{1}+\varphi_{2})}\vert x_{1\downarrow}, x_{2\downarrow} \rangle}{\sqrt{2}} \, ,
\end{align}
the external degrees remains entangled. The same behavior is found by detecting any one component. For more than one dimension, detection of a single atomic component does not define the state of the other components unequivocally and instead leaves them in an entangled state. The atomic components mask one another so they cannot be determined unambiguously.

%%%%%%%%%%%%%%%%%%%%%%%%%%%%%%%%%%%%%%%%%%%%%%%%%%%%%%%%%%%%%%%%%%%%%%
\section{Quantum open system: driven-dissipative quantum phase transition}\label{Sec:open_system}
%%%%%%%%%%%%%%%%%%%%%%%%%%%%%%%%%%%%%%%%%%%%%%%%%%%%%%%%%%%%%%%%%%%%%%

The importance of this superposition becomes apparent when the system is coupled to an external environment that is continuously probing the field. If the system is primarily coupled to the environment through one component, \textit{e.g.} photons that leak through a partial transmitting cavity wall, state-masking allows for entanglement to remain for long times in this dissipative setting. In the following the model is extended to include dissipation in the form of photons leaving the cavity at a rate $\kappa$, the cavity linewidth. An open system configuration also presents advantages as the system can be manipulated externally and the information regarding the state of the system can be obtained in real time by detecting the outgoing photons. In order to explore the effects of atomic state masking an external coherent field is also considered. This external field is used to drive the cavity out of the trivial ground state and allow for a steady-state that displays a non-zero photon number. With the inclusion of drive and dissipation, the master equation for the system density operator is
\begin{equation}\label{eq:master}
\dot{\rho} = \frac{1}{i \hbar} \left[\hat{\mathcal{H}} + \hat{\mathcal{H}}_{D}, \rho \right] + \kappa (2\hat{a} \rho \hat{a}^{\dagger} - \rho \hat{a}^{\dagger}\hat{a} - \hat{a}^{\dagger} \hat{a} \rho ) \, ,
\end{equation}
where $\hat{\mathcal{H}}$ is given in Eq.~(\ref{eq:on-resonance}) and $\hat{\mathcal{H}}_{D}$ accounts for a coherent field of amplitude $\epsilon$ and frequency $\omega_{c}$ acting over the cavity mode
\begin{equation}
\hat{\mathcal{H}}_{D} = \hbar \epsilon(e^{i\omega_{c}t}\hat{a}+e^{-i\omega_{c}t}\hat{a}^{\dagger}) \, .
\end{equation}

The driving field presents the system with another appealing feature as it introduces a competition between coupling strength and coherent field amplitude. When the former dominates, states that minimize fluctuations in photon number are expected, but as the latter dominates the states acquire a preferred phase to minimize their energy cost. This competition is known to cause a system described by Eq.~(\ref{eq:master}) to undergo a phase transition when only internal degrees of freedom are considered. These transitions occurs at a critical driving field amplitude~\cite{Carmichael_2015,Fink_2017}
\begin{equation}\label{eq:critical_drive}
\epsilon_{\text{crit}} = {\textstyle{\frac12}} \Omega  \, ,
\end{equation}
and is reflected on the lowest quasi-energy steady-state. The lowest quasi-energy state displays zero photon-number expectation below the critical drive amplitude and a growing value above it~\cite{Carmichael_2015}. Such response by the cavity mode is a symptom of ordering on the underlying system. While below a critical amplitude the atom radiates to cancel the incoming field, above the critical amplitude atom and cavity field organize to radiate in one of two possible configurations ruled by the dressed state branches~\cite{Alsing_1991}. The configurations correspond to a system driven up one of the two dressed state branches. Such ordering is monitored by detecting the cavity output channel~\cite{Fink_2017}. 

It is in this driving process where the center-of-mass affects the known dynamics. The ordered phase is accompanied by a potential that localizes the center-of-mass, creating multiple paths for the state to be driven up that are not just dependent on the dressed-state branches. Since the expectation values of cavity-mode operators are not able to distinguish between the paths this leads to degenerate subspaces where coherence survives for long times. Adopting the language of Sec.~\ref{Sec:entanglement}, when atomic components mask one another the field scattered from $\vert +, x_{\uparrow} \rangle$ must be equal to that from $\vert -, x_{\downarrow} \rangle$ under Eq.~(\ref{eq:master}). The first question to ask is whether the existence of these two degenerate and incompatible states prevent the system from reaching an ordered phase? In Section~\ref{Sec:ordering} this question is answered in the negative before the conditions for the underlying coherences to appear are presented.

%%%%%%%%%%%%%%%%%%%%%%%%%%%%%%%%%%%%%%
\subsection*{Restricted momentum space}
%%%%%%%%%%%%%%%%%%%%%%%%%%%%%%%%%%%%%%

To study ordering of the open system it is sufficient to restrict ourselves to one spatial dimension. This is possible since the system is being monitored by the photon field, so the pairing between the electronic state and one momentum component is enough to mask the atomic state. Such restriction is not possible using the closed-system examples above where photon number and two-state energy exhibit perfect correlation and two external degrees were needed: one to mask the internal state and another to probe the system. 

It is also convenient to restrict the momentum space to three values only, $ k = 0 , \pm k_{m}$~\cite{Nagy_2010}, since we are interested in monitoring the phases of the system. The kinetic energy within this restricted space is 
\begin{eqnarray}\label{eq:mom_rest}
\frac{1}{2M}\hat{p}_{m}^{2} &=& \frac{\hbar \omega_{\text{r}}}{2} \left(\vert k_{m} \rangle \langle k_{m} \vert - \vert 0 \rangle \langle 0 \vert \right)  \, , 
\end{eqnarray}
taking one of two values separated by the atomic recoil frequency. The operator $\vert k_{m} \rangle \langle k_{m} \vert $ projects over the unit momentum subspace, expanding along positive and negative values. The zero and one momentum states are coupled through the dipole coupling strength
\begin{eqnarray}\label{eq:pos_rest}
2\Omega \cos k_{m}x_{m} &=&  \Omega \left(\vert k_{m} \rangle \langle 0 \vert + \vert 0 \rangle \langle k_{m} \vert \right) \, .
\end{eqnarray}
The external dynamics can then be described by the two-state operators:
\begin{align}
\hat{\mathcal{J}}_{3} &= \vert k_{m} \rangle \langle k_{m} \vert - \vert 0 \rangle \langle 0 \vert \, , \\
\hat{\mathcal{J}}_{+} &= \hat{\mathcal{J}}_{-}^{\dagger} = \vert k_{m} \rangle \langle 0 \vert  \, ,
\end{align}
The interaction Hamiltonian in this restricted space takes the form 
\begin{equation}
\hat{\mathcal{H}} = \frac{\hbar \omega_{\text{r}}}{2} \hat{\mathcal{J}}_{3} + \hbar \omega_{c} \left( \hat{a}^{\dagger} \hat{a} +  \hat{\sigma}_{+}\hat{\sigma}_{-} \right) + \frac{\hbar \Omega}{2} \left(\hat{\mathcal{J}}_{+}+\hat{\mathcal{J}}_{-}\right) \left( \hat{a} \hat{\sigma}_{+} + \hat{a}^{\dagger} \hat{\sigma}_{-} \right)  \, 
\end{equation}
and exhibits both Jaynes-Cummings and Rabi interactions. The former is between internal states and free cavity field and the latter is between center-of-mass states interacting with the dressed field. By restricting the momentum space to three levels, a geometrical view of the dynamics is made possible. When the coupling strength dominates, the eigenstates of $\hat{\mathcal{H}}$ display strong correlations among momentum components that can be interpreted as localized states. These external states point towards the equator of the Bloch sphere with a phase determined by the operator
\begin{equation}\label{eq:localization_close}
\hat{\mathcal{J}}_{1} = \hat{\mathcal{J}}_{+} + \hat{\mathcal{J}}_{-} \, ,
\end{equation}
contrasting the case where recoil energy dominates and the states point towards the poles of the Bloch sphere. Finally, notice that parity chains are now organized by the conserved operator
\begin{equation}
\hat{\mathcal{O}} = \hat{\sigma}_{3} \hat{\mathcal{J}}_{3} \, . \label{eq:restricted_parity}
\end{equation}

%%%%%%%%%%%%%%%%%%%%%%%%%%%%%%%%%%%%%%%%%%%%%%%%%%%%%%%%%%%%%%%%%%%%%%%%%%%%%%%%
\section{Mean-field approximation: incompatible solutions}\label{Sec:mean-field}
%%%%%%%%%%%%%%%%%%%%%%%%%%%%%%%%%%%%%%%%%%%%%%%%%%%%%%%%%%%%%%%%%%%%%%%%%%%%%%%

A first hint towards ordering of the system is found in the steady-state solutions of the mean-field equations. Mean-field equations are obtained from expectations in the Heisenberg equations of motion under the assumption that expectations of operator products factorize. For a system evolving under Eq.~(\ref{eq:master}) the internal dynamical variables: field amplitude $\alpha \equiv\langle \hat{a} \rangle$, atomic polarization $\beta\equiv2\langle \hat{\sigma}_-\rangle$, and population $\zeta\equiv\langle \hat{\sigma}_3\rangle$, satisfy
\begin{eqnarray}
\frac{d\alpha}{dt}&=&-\kappa\alpha-i\textstyle{\frac14}\Omega X_{k} \beta-i\epsilon,
\label{eqn:mean-field_alpha}\\
\frac{d\beta}{dt}&=& i\Omega X_{k} \alpha\zeta,
\label{eqn:mean-field_beta}\\
\frac{d\zeta}{dt}&=& -i\textstyle{\frac12}\Omega X_{k} \left(\alpha\beta^*-\alpha^*\beta\right),
\label{eqn:mean-field_zeta} 
\end{eqnarray}
in a frame oscillating with the cavity frequency $\omega_{c}$. In turn, the external variables: localization $X_{\text{k}} = \langle \hat{\mathcal{J}}_{+} + \hat{\mathcal{J}}_{-} \rangle$, $Y_{\text{k}} = i\langle \hat{\mathcal{J}}_{+} - \hat{\mathcal{J}}_{-} \rangle$, and momentum difference $Z_{\text{k}} = \langle \hat{\mathcal{J}}_{3} \rangle$, satisfy
\begin{eqnarray}
\frac{d X_{\text{k}}}{dt}&=& \omega_{\text{r}} Y_{\text{k}} \, ,
\label{eqn:mean-field_Jbeta}\\
\frac{d Y_{\text{k}}}{dt}&=& -\omega_{\text{r}} X_{\text{k}} + \textstyle{\frac12}\Omega Z_{\text{k}} (\alpha \beta^{*} + \alpha^{*} \beta) \, ,
\label{eqn:mean-field_Jbeta2}\\
\frac{d Z_{\text{k}}}{dt}&=& - \textstyle{\frac12}\Omega Y_{\text{k}} \left(\alpha \beta^{*} + \alpha^{*} \beta \right) \, .
\label{eqn:mean-field_Jzeta}
\end{eqnarray}

Working from Eq.~(\ref{eqn:mean-field_beta}) two steady-state solutions are readily found: (i) a trivial solution where cavity and atom remain decoupled since the center-of-mass is de-localized $(X_{k} = 0)$; (ii) a non-trivial solution where vanishing population $(\zeta = 0$) gives way to a unit polarization
\begin{equation}
\beta = e^{i \varphi} \, . \label{eqn:mean-field_polarization}
\end{equation}
through the conserved quantity $\vert \beta \vert^{2} + \zeta^{2} = 1$. A third solution where an empty cavity ($\alpha = 0$) is caused by the destructive interference of radiated and driving fields is inhibited for free atoms with finite mass (non-zero recoil energy). 

The non-trivial solution appears to encompass this third behavior as it describes a cavity driven by the radiated field of a polarized atom interfering with the external field. By inserting Eq.~(\ref{eqn:mean-field_polarization}) into Eq.~(\ref{eqn:mean-field_alpha}) under the steady-state condition, the field quadratures 
\begin{align}
\alpha_{R} &= {\textstyle{\frac12}}\frac{\epsilon_{\text{crit}}}{\kappa} X_{\text{k}} \sin \varphi \, , \label{eqn:mean-field_quadratureR} \\
\alpha_{I} &= -{\textstyle{\frac12}}\frac{\epsilon_{\text{crit}}}{\kappa} X_{\text{k}} \cos \varphi - \frac{\epsilon}{\kappa} \label{eqn:mean-field_quadratureI}\, .
\end{align}
are found with $2\epsilon_{\text{crit}} = \Omega$ as defined above. The superposition of driving and radiated field creates a potential that can localize the atom. The localization is obtained from Eq.~(\ref{eqn:mean-field_zeta}) and Eqs.~(\ref{eqn:mean-field_polarization})-(\ref{eqn:mean-field_quadratureI}), to give
\begin{equation}
X_{\text{k}} = \frac{2 \epsilon}{\epsilon_{\text{crit}}} \cos \varphi \, , \label{eqn:mean-field_localization}
\end{equation}
where the phase of the dipole $\varphi$ is given by solutions of the transcendental equation
\begin{equation}
\cos^{4}\varphi - \left( \frac{\omega_{\text{r}}\kappa}{2\epsilon^{2}}+\frac{\epsilon_{\text{crit}}^{2}}{\epsilon^{2}} + 1 \right)  \cos^{2}\varphi + \frac{\epsilon_{\text{crit}}^{2}}{\epsilon^{2}}=0 \, . \label{eqn:mean-field_transcendental}
\end{equation}
This is obtained from Eqs.~(\ref{eqn:mean-field_Jbeta}) and~(\ref{eqn:mean-field_localization}) using the conservation law ${X}_{k}^{2} + {Y}_{k}^{2} + {Z}_{k}^{2} = 1$. It is worth comparing this solution to those where only inner degrees are considered (limit of infinite mass). In that case $X_{\text{k}}$ is a constant and the phase of the dipole is locked to that of the field thus lowering the order of the transcendental equation (see Eq.~(42) on Ref.~\cite{Gutierrez_2018}).

\begin{figure}[h]
\begin{center}
\includegraphics[width=.4\linewidth]{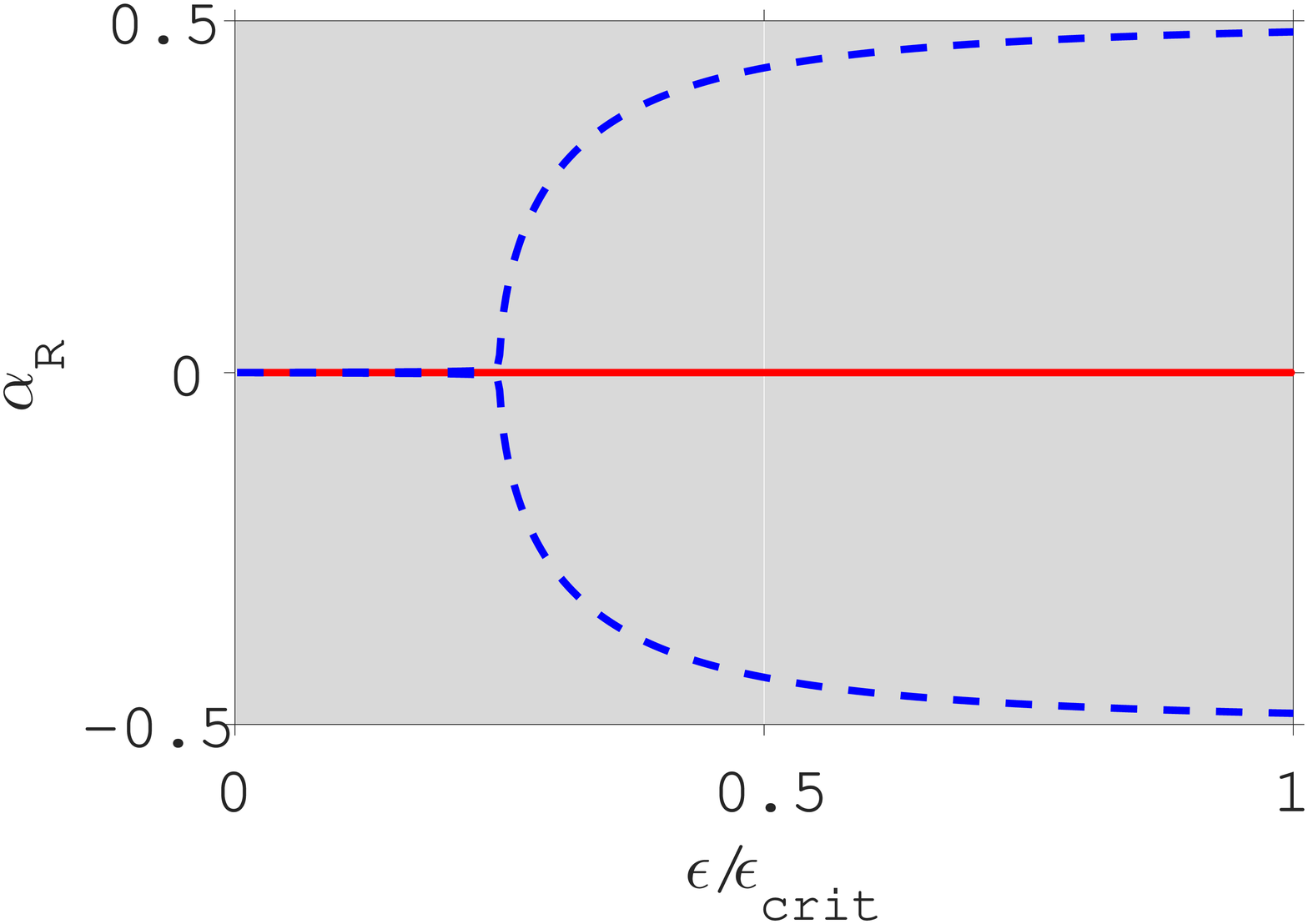} \includegraphics[width=.4\linewidth]{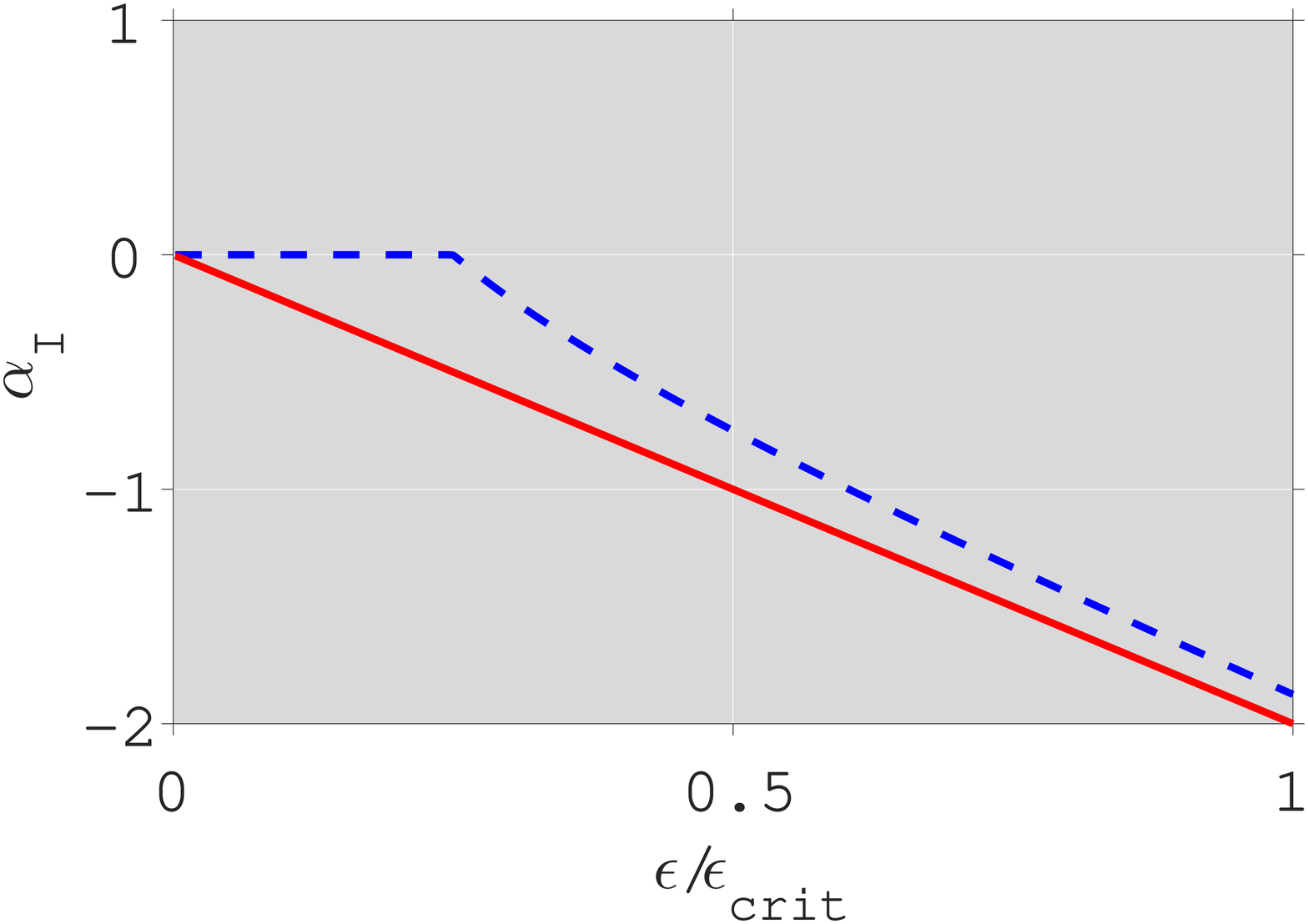}
\caption{Steady-state field quadratures obtained from mean-field Eqs.~(\ref{eqn:mean-field_alpha})-(\ref{eqn:mean-field_Jzeta}) for $\omega_{\text{r}}=.25 \kappa$, $\Omega = 20 \kappa$. Red solid (blue dashed) lines correspond to stable (unstable) states as found from a linear stability analysis. Field quadratures are scaled by $\Omega/\kappa$. }\label{figure_meanfield}
\end{center}
\end{figure}

In Figure~\ref{figure_meanfield} the steady-state quadratures of the cavity field are plotted. A linear stability analysis has been used to distinguish between locally stable solutions (plotted here in solid red lines) and unstable ones (dashed blue lines). For this region of the parameter space only the trivial solution is found to be locally stable. With the atom and cavity decoupled under this solution, the cavity field corresponds to that of a driven Lorentzian cavity. Its amplitude grows linearly with the driving field amplitude and its phase is locked to that of the drive. This response is to be compared with the non-trivial one. Non-trivial solutions describe two competing behaviors: nearly vanishing cavity field below a limiting drive amplitude and a growing field above it. As the drive amplitude increases past a limiting value the dressed atom can not radiate a field that is strong enough to cancel the incoming field and spontaneously acquires one of two phases. It is this dressed state polarization that marks the inherent order of the solution~\cite{Alsing_1991}. 

A system evolving under the mean-field equations~(\ref{eqn:mean-field_alpha})-(\ref{eqn:mean-field_Jzeta}) can not reach the ordered state since the non-trivial solution remains unstable throughout. This is attributed to a cavity field that probes an atomic state where the polarization is masked by two possible and incompatible regions of localization, \textit{i.e.}, for the same field value the atom displays two solutions: one where it is localized at the minima and one where it is localized at the maxima of the potential. This is due to its phase satisfying a quartic equation and the dependence of the dipole coupling on a product of internal and external atomic degrees of freedom. These macroscopically incompatible states cause the non-trivial solutions to remain unstable and prevents the system from reaching an ordered phase regardless of the driving field amplitude.

%%%%%%%%%%%%%%%%%%%%%%%%%%%%%%%%%%%%%%%%%%%%%%%%%%%%%%%%%%%%%%%%%%%%%%%
\section{Quantum correlations: ordering of the masked system}\label{Sec:ordering} 
%%%%%%%%%%%%%%%%%%%%%%%%%%%%%%%%%%%%%%%%%%%%%%%%%%%%%%%%%%%%%%%%%%%%%%%%

The cavity-field quadratures in Figure~\ref{figure_meanfield} are rescaled by $\Omega/\kappa$ and the driving field amplitude by $\epsilon_{\text{crit}}$. With the photon-number expectation being bounded by the ratio between incoming and outgoing photon flux $\epsilon^{2}/\kappa^{2}$ and the value required to activate the non-linearity $\Omega^{2}/\kappa^{2}$, the scaling allows us to set up a thermodynamic limit, a limit of high excitation number where fluctuations can be neglected~\cite{Carmichael_2015}. The limit is reached when $\epsilon, \Omega \to \infty$ keeping the ratio $\epsilon/\Omega$ constant. Disorder is seen to remain close to this limit and suggests that, if order is to be found, a description that accounts for the quantum correlations between atomic degrees of freedom has to be considered. To describe the correlations and degeneracies of the system, it is possible to evolve the density matrix under the master equation~(\ref{eq:master}) until it reaches a steady state. The results presented in the following are obtained by performing this evolution numerically~\cite{Ricardo_2019}.

%%%%%%%%%%%%%%%%%%%%%%%%%%%%%%%%%%%%%%%%%%%%%%%%%%%%%%%%
\subsection{Phase-space representation}
%%%%%%%%%%%%%%%%%%%%%%%%%%%%%%%%%%%%%%%%%%%%%%%%%%%%%%%%%

\begin{figure}[h]
\begin{center}
\includegraphics[width=.4\linewidth]{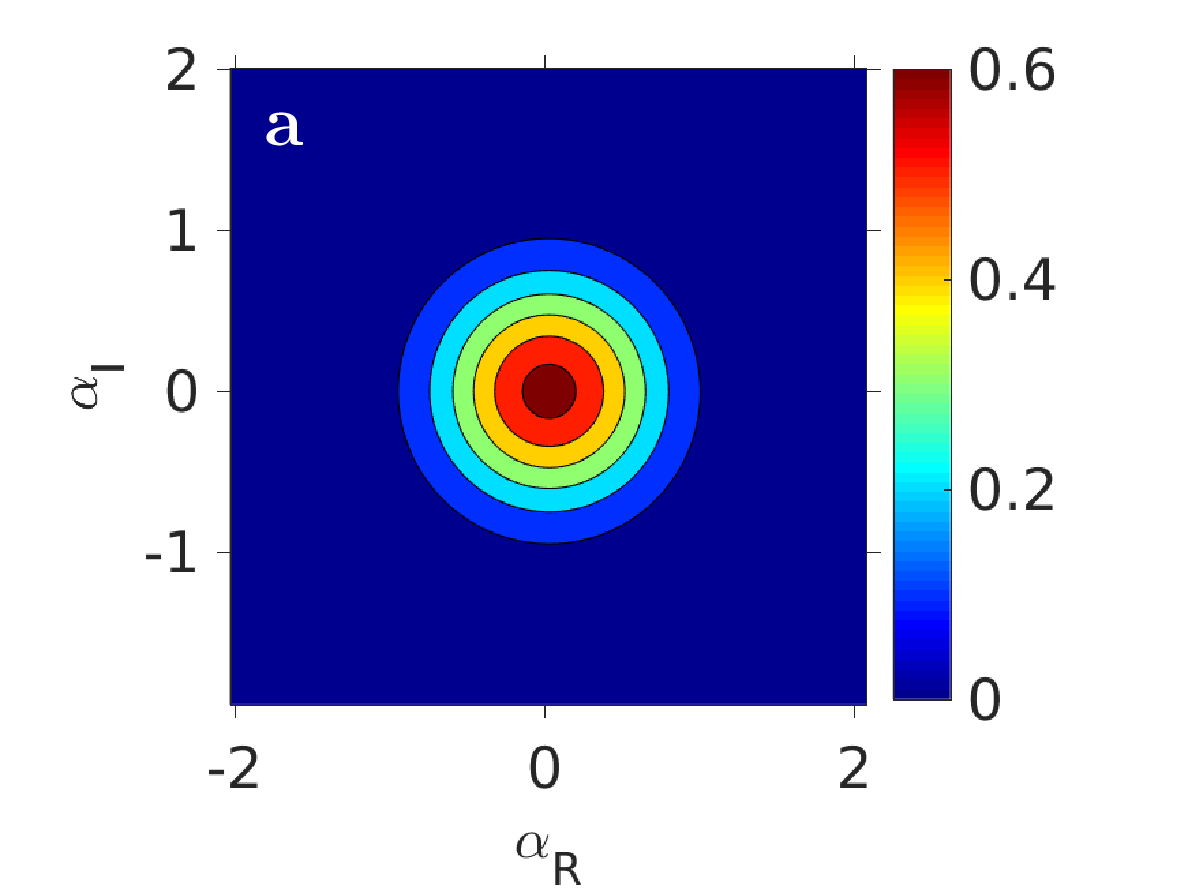} \includegraphics[width=.4\linewidth]{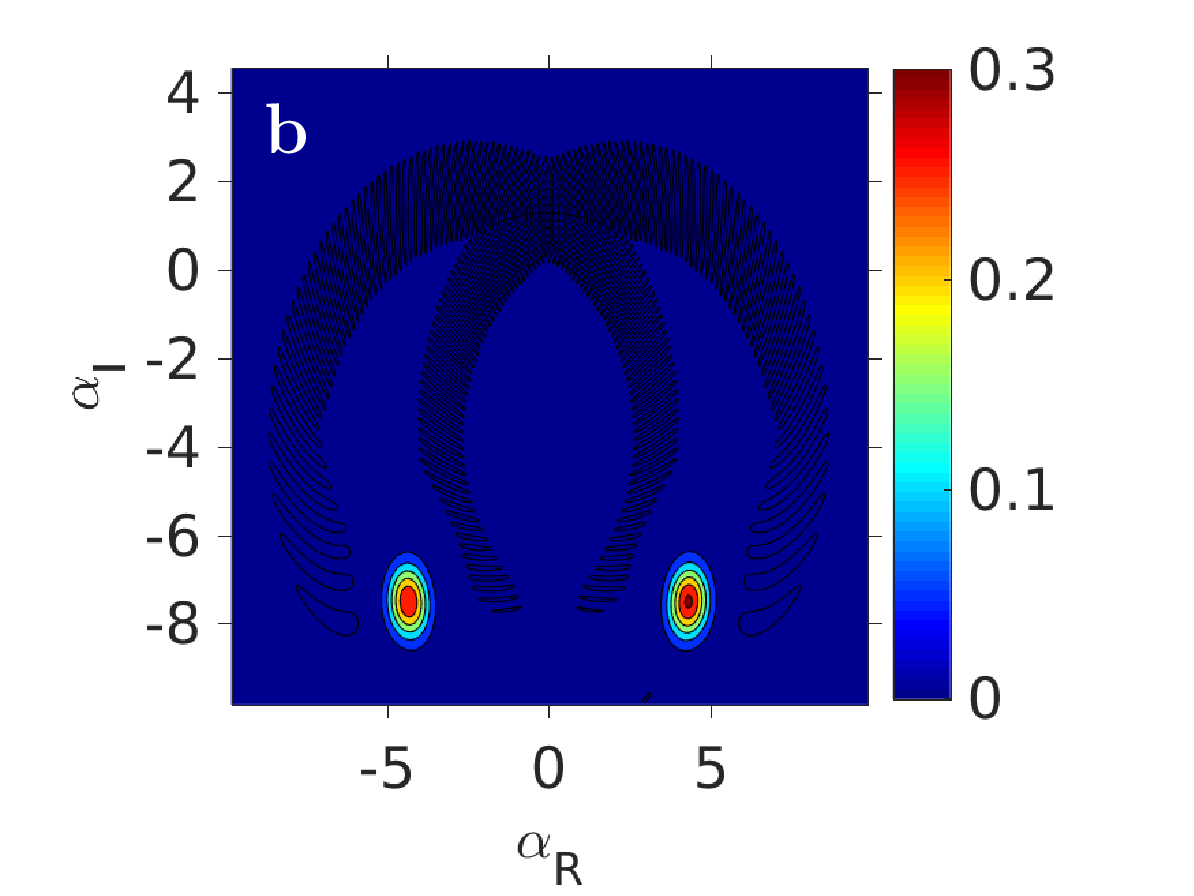}
\caption{Steady-state Wigner distribution for $\omega_{\text{r}}=.25 \kappa$, $\Omega = 20 \kappa$ and driving fields amplitudes $\epsilon = \epsilon_{\text{crit}}/8$ in (a) and $\epsilon = \epsilon_{\text{crit}}$ in (b).}\label{figure_wigner}
\end{center}
\end{figure}

The Wigner distributions of the field in the steady-state are displayed in Figure~\ref{figure_wigner}. The distributions are obtained from two different drive amplitudes (above and below a limiting drive amplitude $\epsilon = \Omega/4$) chosen from mean-field solutions and the driven Jaynes-Cummings model results as a guide~\cite{Gutierrez_2018b}. Frame (a) is reserved for the case of $\epsilon = \Omega /16$, below the limiting field amplitude, where disorder is to be expected. The steady-state is characterized by a single-peaked Wigner distribution centered around the vacuum state. A nearly empty cavity represents a system deep into the photon blockade regime where the non-linear spectrum of the Jaynes-Cummings model causes the absorption of one excitation to shift the mode out of resonance. Frame (b), by comparison, is reserved for a driving field $\epsilon = \Omega/2$ well above the limiting amplitude ($\epsilon = \Omega/4$). The steady-state displays a double-peaked distribution that signals optical bimodality caused by the symmetry breaking and underlying order of the system. The two peaks represent the transmission lines of the system, metastable states displaying high photon-number expectation ($\bar{n}$) that are related to the upper and lower dressed-state branches. This can be seen from the quasi-energy spectrum~\cite{Gutierrez_2018b}. When the drive is increased past the limiting value, the non-linear spectrum gives way to a continuous spectrum where the system is driven up a quasi-harmonic energy ladder ($\sqrt{\bar{n}-1}-\sqrt{\bar{n}} \simeq 0$). Unlike the driven Jaynes-Cummings model, each peak contains a superposition of both dressed-state branches masked by the appropriate atomic localization.

\begin{figure}[h]
\begin{center}
\includegraphics[width=.4\linewidth]{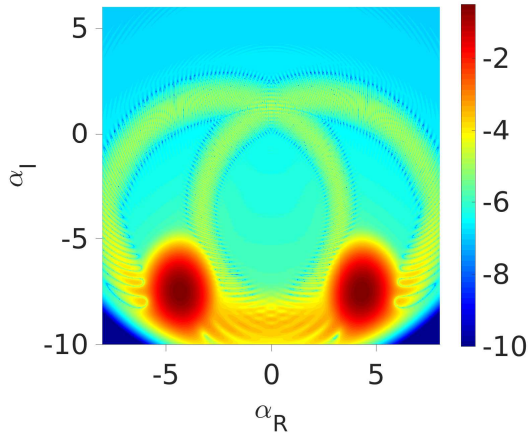}
\caption{Absolute value of the Wigner distribution in the ordered regime using a logarithm scale ($\log \vert W(\alpha, \alpha^{*}) \vert $). The same parameters as Figure~\ref{figure_wigner_2}(b) are taken.}\label{figure_wigner_2}
\end{center}
\end{figure}

The transmission lines are connected, if only weakly, by quantum fluctuations. The loss of a photon allows for transitions from one dressed-state branch to the other through~\cite{Carmichael_2015}
$$ \sum_{n^{\prime}}\langle +,n^{\prime}, \mathcal{J}_{1} \vert \hat{a} \vert -,n, \mathcal{J}_{1} \rangle = \frac{\sqrt{n}-\sqrt{n-1}}{2}$$
where $\vert \mathcal{J}_{1} \rangle$ is an eigenstate of the external operator $\hat{\mathcal{J}}_{1}$ [defined in Eq.~(\ref{eq:localization_close})]. This is represented in Fig.~\ref{figure_wigner}b by two dark paths that connect the peaks. The Wigner distribution takes negative values along these paths. To better appreciate the strength of the fluctuations, the WIgner distribution is plotted using a logarithm scale in Figure~\ref{figure_wigner_2}. As the drive amplitude is ramped up and the system delves deeper into the ordered regime, the peaks grow further apart and the paths become less prominent.

The change in the modality of the system signals a dissipative quantum phase transition on the system~\cite{Bonifacio_1978}, which, up to this point, resembles the breakdown of the photon blockade in the driven Jaynes-Cummings model. In the present case, however, the transition is lost under the mean-field approximation and stabilizes once quantum correlations are taken into account. This is attributed to the atomic state-masking and is one of the main results from this work. The mean-field results are unstable since each transmission line refers to a subspace, a region where a superposition of classically incompatible states is possible. Quantum correlations are required to describe the co-existence of these states.

In fact, the existence of these subspaces was already implied by the numerical method used to evolve the density matrix. The Wigner distributions plotted in Figure~\ref{figure_wigner} are obtained for a system that is initially prepared in the state
\begin{equation}\label{eq:initial_state}
\vert \psi (t_{0}) \rangle = \vert g, n=0, \mathcal{J}_{3}=  +1 \rangle  \, .
\end{equation}
This is an eigenstate of the restricted parity operator $\hat{\mathcal{O}}$ with eigenvalue one [see Eq.~(\ref{eq:restricted_parity})]. Information regarding the initial state is usually not important in driven-dissipative systems as the master equation drives the system to a particular distribution independent of the initial conditions at long times, under standard circumstances~\cite{Minganti_2018}. But with the conservation of the parity branches under the master equation this is not necessarily the case. A state prepared in
\begin{equation}
\vert \psi (t_{0}) \rangle = \vert g, n=0, \mathcal{J}_{3}=  -1 \rangle  \, .
\end{equation}
is an eigenstate of the parity operator with eigenvalue minus one. The evolution of this state leads to a Wigner distribution that is virtually indistinguishable from the one plotted above. But the two states live in different spaces, and the distributions can not overlap. They only lead to similar distributions because the cavity field is not able to distinguish between the parities. As mentioned above, this can be exploited to leave the system in a degenerate subspace where coherences can be preserved.  

%%%%%%%%%%%%%%%%%%%%%%%%%%%%%%%%%%%%%%%%%%%%%%%%%%%%%%%%
\subsection{Quantum trajectory theory}
%%%%%%%%%%%%%%%%%%%%%%%%%%%%%%%%%%%%%%%%%%%%%%%%%%%%%%%%%

While information of the atomic state remains hidden from the field, an external observer monitoring the outgoing field can infer the state of the system conditioned to a particular measurement record using quantum trajectory theory~\cite{Carmichael_2008}. The scheme selected to monitor the outgoing field must be able to detect its phase. For it is the phase, rather than the photon-number expectation, that differentiates between the two peaks in the Wigner distribution. Heterodyne detection is a good candidate for this measurement, where current records are obtained by mixing the output signal with an intense coherent field  that provides a phase reference. The resulting field is then detected and the outgoing field quadratures can be measured from this detection. This process can be simulated by a stochastic Schr\"odinger equation when the local oscillator amplitude is strong. The evolution of the system state vector $\vert \psi_{\text{\tiny{REC}}}\rangle$ conditioned to a particular measurement record satisfies
\begin{equation}
d \vert \psi_{\text{\tiny{REC}}} \rangle = \left[ \frac{1}{i \hbar} \left( \hat{\mathcal{H}} -i\hbar \kappa \hat{a}^{\dagger} \hat{a} \right) dt  + \sqrt{2\kappa} \hat{a} dq \right] \vert \psi_{\text{\tiny{REC}}} \rangle \, ,
\end{equation}
where $dq$ represents the heterodyne counts accumulated in the time interval $t$ and $t+ dt$~\cite{Carmichael_2008}. The counts depends on the conditional state through the equation
\begin{equation}
dq =  \frac{\langle \psi_{\text{\tiny{REC}}} \vert \sqrt{2\kappa} \hat{a}^{\dagger} \vert \psi_{\text{\tiny{REC}}} \rangle}{\langle \psi_{\text{\tiny{REC}}} \vert \psi_{\text{\tiny{REC}}} \rangle} dt + dZ \, .
\end{equation}
where the complex Wiener increment $dZ$ accounts for the shot noise of the local oscillator. The filtered heterodyne current is constructed from these counts using 
\begin{equation}\label{eq:current_heterodyne}
dI = -\kappa_{\text{\tiny{D}}} \left[ I dt - \frac{dq}{\sqrt{\kappa}} \right]  \, , 
\end{equation}
where $\kappa_{\text{\tiny{D}}}$ is the linewidth of the detector. 

A generic heterodyne record is composed of the real and imaginary components of the current. For the system at hand the imaginary component is centered around a given value, while the real current transitions among two values corresponding to each peak of the Wigner distribution [see Fig.~\ref{figure_wigner}(b) above]. These transitions occur at random times. They represent a change in the ordering of the system and are driven by fluctuations. A toy model describing the conditioned states helps to clarify how the atomic state is conditioned to the values found in a particular heterodyne record. The model is based on intuition gathered from the analytic solutions of the driven Jaynes-Cummings model~\cite{Gutierrez_2018b}. When the real part of the photo-current displays a positive value, the conditioned state can be expanded as a superposition of the states
\begin{eqnarray}
\vert \psi^{+}_{\text{\tiny{REC}}} \rangle = \sum_{n} c_{n} (e^{i\varphi_{u}} \vert + , n , \mathcal{J}_{1}=1 \rangle + e^{i\varphi_{d}}\vert - , n , \mathcal{J}_{1}=-1 \rangle) \, ,
\end{eqnarray}
where phases $\varphi_{u,d}$ depend on the parity of the initial states from Eq.~(\ref{eq:localization_mask}). Notice the external component is described in the localization basis rather than momentum population basis. This masked atomic state is inferred from measurements of a scattered field that carries an phase $+\phi$. This undisclosed phase is related to the relative phase between coupling and driving strengths since applying the interaction Hamiltonian over these states leads to a dipole coupling of $+\hbar\Omega/2$. In the same manner, when the real part of the photo-current takes a negative value, the conditioned state is
\begin{equation}
\vert \psi^{-}_{\text{\tiny{REC}}} \rangle = \sum_{n} c^{\prime}_{n} (e^{i\varphi_{u}} \vert + , n , \mathcal{J}_{1}=-1 \rangle + e^{i\varphi_{d}}\vert - , n , \mathcal{J}_{1}=1 \rangle)  \, ,
\end{equation}
such that the scattered field carries a phase $-\phi$ (related to $-\hbar\omega/2$ coupling). A system conditioned to a heterodyne record remains entangled in the atomic components.

\begin{figure}[h]
\begin{center}
\includegraphics[width=.8\linewidth]{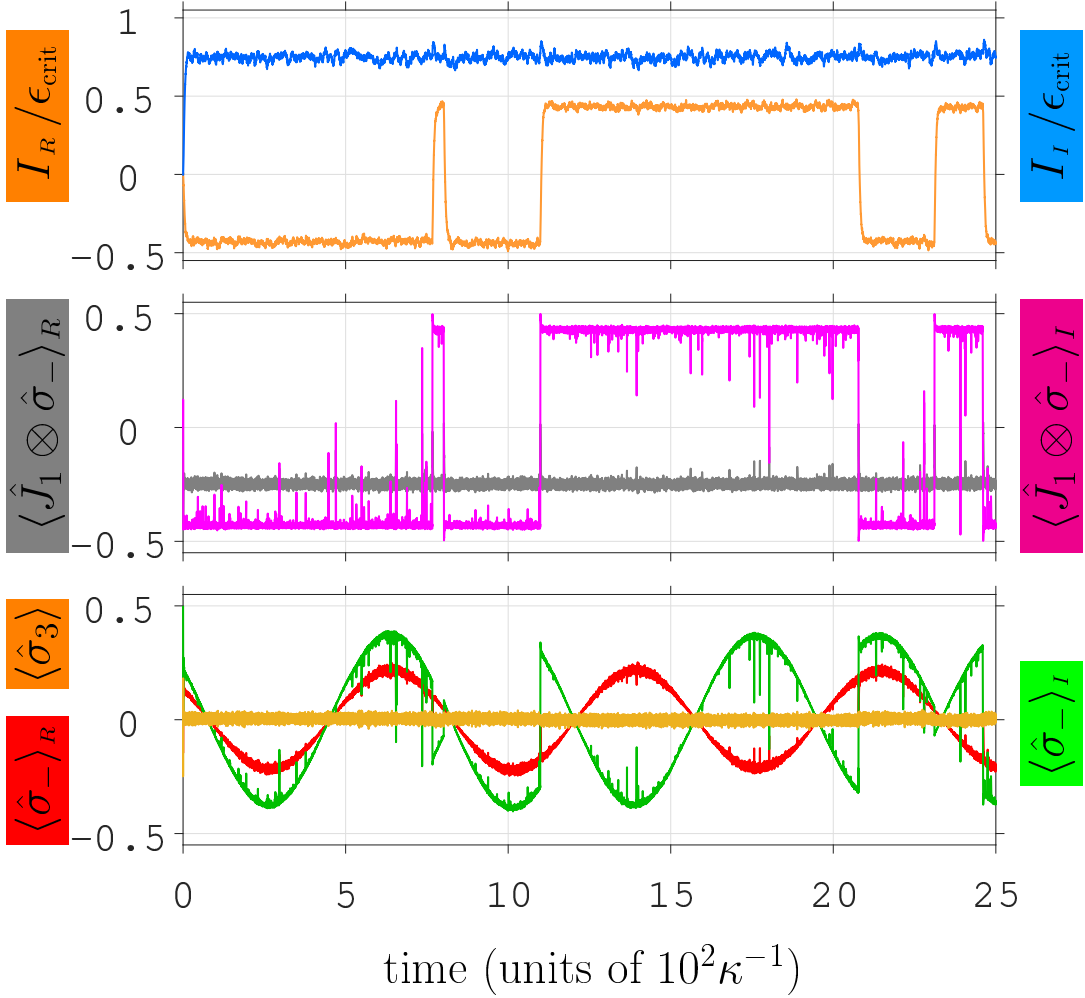} 
\caption{Sample quantum trajectory for a system prepared in state~(\ref{eq:trajectory_start}) with $\omega_{\text{r}}=.25 \kappa$, $\Omega = 20 \kappa$, $\epsilon = \epsilon_{\text{crit}}$, and $\kappa_{\text{\tiny{D}}}=0.25\kappa$. The time is measured in units of inverse cavity linewidths $\kappa^{-1}$. Top frame is reserved for the field quadratures obtained from Eq.~(\ref{eq:current_heterodyne}), middle frame for atomic correlations, and bottom frame for atomic internal state.}\label{figure_heterodyne}
\end{center}
\end{figure}

It is now important to show under which conditions these entangled states can be chosen to display coherence in the atomic variables. When the system is initially prepared in a single parity chain, the expectations $\langle \hat{\sigma}_{-} \rangle_{{\text{\tiny{REC}}}}$ and $\langle \hat{\mathcal{J}}_{+}+\hat{\mathcal{J}}_{-} \rangle{\text{\tiny{REC}}}$ are identically zero. This can be shown by direct substitution with appropriate $\phi_{u,d}$ phases. This is caused by the partial trace over an atomic component destroying the coherence of the process. Yet, when starting from a superposition of the two chains the expectations display slow oscillations (compared to the Rabi and recoil frequencies) determined by the beating frequency of both parities. A sample quantum trajectory is shown using a frame sequence in Fig.~\ref{figure_heterodyne} to reveal these oscillations. The sample trajectory is taken for the initial state
\begin{equation}\label{eq:trajectory_start}
\vert \psi_{\text{\tiny{REC}}}(t_{0}) \rangle = \frac{ \vert g,0,-1\rangle + i \vert e , 0, -1\rangle }{\sqrt{2}}
\end{equation}
with equal components over both parity chains. The top frame shows the simulated photo-current record. It exhibits two locally stable solutions where the phase of the field remains trapped for many cavity lifetimes until fluctuations drive it towards the other. For this record, the first three transitions occur around 700, 800, and 1100 cavity lifetimes. The measured field signals a particular ordering of the system as suggested by the expectation $\langle (\hat{\mathcal{\mathcal{J}}}_{+}+\hat{\mathcal{J}}_{-}) \hat{\sigma}_{-} \rangle$ portrayed in the second frame. This operator expectation also takes one of two possible values predicted by the toy model. Knowledge of the outgoing field conditions the atom into a particular subspace to evolve in. When the measured field transitions between stable values so does the atomic expectation. The coherent dynamics encountered within each subspace remain hidden from the field. It isn't until expectations $\langle \hat{\sigma}_{-} \rangle$ or $\langle \hat{\mathcal{J}}_{+}+\hat{\mathcal{J}}_{-} \rangle$ are computed that the coherence is revealed. The polarization expectation value $\langle \hat{\sigma}_{-} \rangle$ is portrayed in the bottom frame. It displays the desired slow oscillations at an undetermined frequency. The expectation value is still conditioned to the state of the field as it displays abrupt changes on its imaginary component that accompany the changes on the phase of the field. It, however, displays a long-lived coherence due to the existence of the subspaces. 

The imaginary part of the potential changes abruptly as the field amplitude transitions between the two possible phases. In order for this oscillation to be consistent with the correlations between polarization and localization the latter must display a continuous oscillation. The atomic motion is then described by a slow coherent oscillation from one antinode of the potential to another, with the effective coupling strength that correlates photon number and electronic energies to follow this behavior. This perspective helps explain the low frequency of the oscillations. They are caused by the recoil energy that does not allow for the external states to settle at the equator of the Bloch sphere but instead to perform harmonic motion around it. For zero recoil energy, internal and external fields are uncorrelated and the oscillations are lost. The frequency is then determined by an interplay of recoil frequency and coupling strength. The amplitude of the oscillations, however, is determined by the relative population between parity chains. This is as expected, since the lowering operator transfers states from one parity chain to the other.

%%%%%%%%%%%%%%%%%%%%%%%%%%%%%%%%%%%%%%%%%
\section{Conclusions}\label{Sec:conclusion} 
%%%%%%%%%%%%%%%%%%%%%%%%%%%%%%%%%%%%%%%%%%

The model presented here describes an idealized scenario where a two-state system interacts with a standing-wave cavity mode. The interaction is dipolar in character and describes the exchanges of excitation and momentum that correlate the internal and external degrees of freedom of this composite quantum system. These correlations are commonly exploited to probe the underlying global system. As examples, an atom traversing a cavity has been used to measure the photon number expectation of the cavity mode~\cite{Guerlin_2007,Deleglise_2008} or the spatial distribution of the mode~\cite{Guthohrlein_2001,Steiner_2013}. The detection is not restricted to the atomic variables; the quadratures of a scattered field have been used to probe phase transitions in optical systems and infer the global state~\cite{Baumann_2010,Fink_2017}. In these three examples the measured component provides a good probe for the system, taking values from which the global state of the system could be inferred up to a good approximation.  

I have taken a step in a different direction, where, by considering both internal and external degrees of freedom, the emerging quantum correlations allow for multiple states to display the same value when probed. The selected probe is not a bad one because of fluctuations, but because symmetries of the model allow for degeneracies that involve correlations among several degrees of freedom. The atomic components mask one another so each one cannot be determined unambiguously by, for instance, measuring the scattered field. 

This feature is shown to have striking consequences when the description is extended to include a coherent driving field and the effect of a surrounding electromagnetic environment that constantly measures the state of the system. Here, the competition between driving field amplitude and coupling strength causes the system to undergo a dissipative quantum phase transition but, due to atomic masking, the ordering above a critical point is only seen when quantum correlations are taken into account. This causes a departure between mean-field results and the thermodynamic limit found under the quantum evolution. The latter leads to a double peaked Wigner distribution of the field above the critical point. As the thermodynamic limit is reached and the effect of fluctuations neglected, the paths connecting both peaks dissapear and the system displays optical bistability. In contrast, the mean-field solutions describe two macroscopically incompatible states that inhibit any form of optical bistability. This raises an interesting point when considering the analogy between quantum phase transitions in optical systems and those of closed systems in thermal equilibrium~\cite{Bonifacio_1978}. Under this analogy the steady-states play the role of the ground state and the mean-field equations provide a blueprint for the phase diagram. However, as shown here, the mean-field solutions fail to announce an organization in the underlying system when quantum correlations survive the effect of dissipation. In this case the organization does not correspond to the standard formation of regular patterns; since atomic degrees of freedom are masked, state superpositions are found within an ordered state. These states are expected to appear in experiments as the control of external and internal degrees of quantum states continues to increase.

Ultimately these correlations can be exploited to store coherence on the system. This atomic coherence can be probed experimentally using other forms of driving. In Ref.~\cite{Domokos_2002a} Domokos, Salzburguer, and Ritsch explore the case where atom, rather than cavity mode, is driven by an external coherent field. The cavity is excited by the photons scattered from the atom and ordering is still expected above a critical drive amplitude. In this case the driving field enters through the side of an open cavity and breaks the parity symmetry [see Eq.~(\ref{eq:parity_z})] due to the possibility of exciting the electronic state without scattering photons into the cavity mode. By breaking the symmetry, it becomes possible to use this driving to probe the atomic components individually. In this sense, an experiment where atomic masking can be detected is feasible using a setting similar to those in Refs.~\cite{Baumann_2010,Leonard_2017} with the addition of a second driving field for the cavity mode. 

%%%%%%%%%%%%%%%%%%%%%%%%%%%%%%%%%%%%%%%%%%%%%%%%%%%%%%%%%%%%%%%%%%%%%%%%%%%%%%%%%%%%%%%%%%%%%%%%%%%%%%%%%%%%%%%%%%%%%%%%%%%%%%%%%%%%%%%%%%%%%%%%%%%%%%%%%%%%%%%%%%%%%%%%%%%%%%%%%%%%%%%%%%%%%%%%%%%%%%%%%%

\section*{Acknowledgments}

I thank A.~Giraldo, Th.~Mavrogordatos, and R.~J\'{a}uregui for insightful discussions and C.~Langlett and S.~Masson for helpful comments on the manuscript. Support by the Institute for Quantum Science and Engineering is acknowledged with gratitude. This work was supported by the Robert A. Welch Foundation Postdoctoral Fellowship (Welch Foundation Grant No. A-1547).


\begin{thebibliography}{1}
\bibitem{Ricardo_2019} The evolution is performed in a truncated photon-number space for the initial states $\vert g, n=0, \mathcal{J}_{3}=  +1 \rangle$ or $\vert g, n=0, \mathcal{J}_{3}=  -1 \rangle$. Representing the ground state of each parity branch.
\end{thebibliography}
\end{document}